# Scalable Gaussian Processes for Predicting the Properties of Inorganic Glasses with Large Datasets

*Suresh Bishnoi[1], R. Ravinder[1], Hargun Singh[1], Hariprasad Kodamana[2,*], N. M. Anoop Krishnan[1,3,*]*

[1]Department of Civil Engineering, Indian Institute of Technology Delhi, Hauz Khas, New Delhi 110016, India
[2]Department of Chemical Engineering, Indian Institute of Technology Delhi, Hauz Khas, New Delhi 110016, India
[3]Department of Materials Science and Engineering, Indian Institute of Technology Delhi, Hauz Khas, New Delhi 110016, India

*Corresponding authors: H. Kodamana (kodamana@iitd.ac.in), N. M. A. Krishnan (krishnan@iitd.ac.in)

**Abstract**
Gaussian process regression (GPR) is a useful technique to predict composition–property relationships in glasses as the method inherently provides the standard deviation of the predictions. However, the technique remains restricted to small datasets due to the substantial computational cost associated with it. Here, using a scalable GPR algorithm, namely, kernel interpolation for scalable structured Gaussian processes (KISS-GP) along with massively scalable GP (MSGP), we develop composition–property models for inorganic glasses based on a large dataset with more than 100,000 glass compositions, 37 components, and nine important properties, namely, density, Young's, shear, and bulk moduli, thermal expansion coefficient, Vickers' hardness, refractive index, glass transition temperature, and liquidus temperature. Finally, to accelerate glass design, the models developed here are shared publicly as part of a package, namely, Python for Glass Genomics (PyGGi, see: http://pyggi.iitd.ac.in).

**Introduction**
Despite the discovery of over 200,000 glass compositions, the knowledge of composition–property relationships remains restricted to a few selected compositions[1,2]. Development of reliable composition–property maps for a large class of glass components is the bottleneck impeding the design of new glass compositions. Machine learning (ML) methods [3–8] have been used to predict properties such as Young's modulus[9,10], liquidus temperature[11], solubility[12], glass transition temperature[4], and dissolution kinetics[5,13]. Recently, we developed deep learning models to predict composition–property maps of inorganic glasses with 37 input components and eight properties, largest thus far[14]. However, most of these studies employ deterministic models in predictions such as neural networks (NN), random forest, or simple regression models. As such, these models cannot provide any information about the reliability of the predictions for any new test data.

Since the ML methods are primarily data-driven predictions, the reliability of the model is highly dependent on the available data. To this extent, Gaussian process regression (GPR)[15], a nonparametric ML model, presents an excellent candidate. GPR employs a probabilistic approach wherein the inference on new data is made by learning the underlying distribution (mean and covariance) of the available data[15]. It has been shown that for small datasets, GPR models are more suitable in comparison to NN models for providing accurate composition–property predictions along with its confidence intervals in oxide glasses[9]. However, for large datasets available in materials science, training the conventional GPR, which has $\mathcal{O}(n^3)$ time and $\mathcal{O}(n^2)$ space complexity for a dataset of size *n*, is computationally prohibitive.

Here, using scalable GPR algorithms, namely, kernel interpolation for scalable structured Gaussian processes (KISS-GP)[16] and massively scalable Gaussian processes (MSGP)[17], we address the challenge of developing reliable GPR models for predicting nine relevant properties, *viz.*, density, Young's, shear, and bulk moduli, thermal expansion coefficient (TEC), Vickers' hardness, refractive index, glass transition temperature ($T_g$), and liquidus temperature of functional glasses based on a large dataset with more than 100,000 glasses and 37 components. These are the largest models developed to predict composition–property relationships in inorganic glasses. We show that KISS-GP, along with MSGP, provides rigorous models for large datasets that are superior to the state-of-the-art deep neural network (DNN) models[14]. Further, the models provide estimates of the uncertainty associated with the predictions, making these models more reliable and robust in comparison to DNN models. Overall, we show that the methodology presented here can be used for developing GPR models for problems with large training datasets. Finally, the models developed here are made available as part of a software package designed for accelerating glass discovery, namely, Python for Glass Genomics (PyGGi, see: http://pyggi.iitd.ac.in).

**Methodology**
**Dataset Preparation**
The raw dataset consisting of nine properties, namely, density, Young's modulus, bulk modulus, shear modulus, Vickers' hardness, glass transition temperature, liquidus temperature, thermal expansion coefficient, and refractive index of inorganic glasses are obtained from the literature and commercial databases such as INTERGLAD Ver. 7. Here we follow a rigorous dataset preparation employed earlier to develop deep learning models for glass property models [14]. The basic steps involved in the methodology are as follows. The duplicate entries in the dataset are removed. The raw dataset consists of glass compositions with approximately 270 components. However, many of these components out of the 270, are present in a few glass compositions only. Such a sparse dataset may lead to a poorly trained model, as enough representative samples may not be present in the training and test set. To overcome this challenge, we employ the least angle regression (LARS) for dimensionality reduction. In particular, the input parameters (that is, glass components) are chosen based on the covariance from the dataset, thereby drastically reducing the glass components while still preserving a reasonable training set for accurate predictions. The final number of glasses consists of oxide components ranging from 32 to 37. The frequency of glasses corresponding to each of the components is provided in the Supplementary material. The final dataset consists of 37 components, namely, $SiO_2$, $B_2O_3$, $Al_2O_3$, $MgO$, $CaO$, $BaO$, $Li_2O$, $Na_2O$, $K_2O$, $Ag_2O$, $Cs_2O$, $Ti_2O$, $BeO$, $NiO$, $CuO$, $ZnO$, $CdO$, $PbO$, $Ga_2O_3$, $Y_2O_3$, $La_2O_3$, $Gd_2O_3$, $Bi_2O_3$, $TiO_2$, $ZrO_2$, $TeO_2$, $P_2O_5$, $V_2O_5$, $Nb_2O_5$, $Ta_2O_5$, $MoO_3$, $WO_3$, $H_2O$, $Sm_2O_3$, $MgF_2$, $PbF_2$, and $PbCl_2$, and nine properties, namely, density, Young's, shear, and bulk moduli, TEC, Vickers' hardness, refractive index, $T_g$, and liquidus temperature.

**Gaussian Process Regression (GPR)**
Gaussian processes (GPs) are models that are capable of modeling datasets in a probabilistic framework. The main advantages of GP models are: (i) its unique ability to model any complex data sets; (ii) estimate the uncertainty associated predictions through posterior variances computations. A GP is a joint distribution of any finite set of random variables that follow Gaussian distributions. As a result, the GPR modeling framework tries to ascribe a distribution over a given set of input (*x*) and output datasets (*y*)[15]. A mean function *m(x)* and a covariance function $k(x,x')$, the two degrees of freedoms that are needed to characterize a GPR fully, are as shown below.

$$y = f(x) + \epsilon; \text{ where } \epsilon \sim N(0, \sigma_\epsilon^2), \text{ and } f \sim GP(m(x), k(x,x')) \quad (1)$$

while the mean function $m(x)$ computes the expected values of output for a given input, the covariance function captures the extent of correlation between function outputs for the given set of inputs as

$$k(x, x') = E[f(x) - m(x)), f(x') - m(x')] \quad (2)$$

In the GP literature, $k(x,x')$ is also termed as the kernel function of the GP. A widely used rationale for the selection of kernel function is that the correlation between any points decreases with an increase in the distance between them. Some popular kernels in the GP literature are

1. Exponential kernel: $k(x, x') = exp \frac{|x-x'|}{l}$ \quad (3)

2. Squared exponential kernel: $k(x, x') = \sigma_f^2 exp\left[-\frac{1}{2}\left(\frac{x-x'}{l}\right)^2\right]$ \quad (4)

where $l$ is termed as the length-scale parameter and $\sigma_f^2$ is termed as the signal variance parameter. In a GPR model, these hyper-parameters can be tuned to model datasets that have an arbitrary correlation. Also, the function $f \sim GP(m(x), k(x,x'))$ is often mean-centered for relaxing the computational complexity.

Suppose, we have a set of test inputs $X_*$ for which we are interested in computing the output predictions. This would warrant sampling as a set of $f_* \triangleq [f(x_{1*}), ..., f(x_{n*})]$, such that $f_* = N(0, K(X_*, X_*))$ with the mean and covariance as

$$m(x) = 0; \quad K(X_*, X_*) = \begin{bmatrix} k(x_1^*, x_1^*) & \cdots & k(x_1^*, x_n^*) \\ \vdots & \ddots & \vdots \\ k(x_n^*, x_1^*) & \cdots & k(x_n^*, x_n^*) \end{bmatrix} \quad (5)$$

By the definition of GP, the new and the previous outputs follow a joint Gaussian distribution as

$$\begin{pmatrix} y \\ f_* \end{pmatrix} \sim N\left(0, \begin{matrix} K(X,X) + \sigma_\epsilon^2 I & K(X, X_*) \\ K(X_*, X) & K(X_*, X_*) \end{matrix}\right) \quad (6)$$

where, $K(X, X)$ is the covariance matrix between all observed inputs, $K(X_*, X_*)$ is the covariance matrix between the newly introduced inputs, $K(X_*, X)$ is the covariance matrix between the new inputs and the observed inputs and $K(X, X_*)$ is the covariance matrix between the observed points and the new inputs, and $I$ is the identity matrix. Now, applying the principles of conditionals, $p(f_*|y)$ can be shown to follow a Normal distribution with:

$$\text{mean}(f_*) = K(X_*, X)(K(X, X) + \sigma_\epsilon^2 I)^{-1} y \quad (7)$$
$$\text{covariance}(f_*) = K(X_*, X_*) - K(X_*, X)(K(X, X) + \sigma_\epsilon^2 I)^{-1} K(X, X_*) \quad (8)$$

Equations (7) and (8) are employed to make new predictions using the GPR.

**Kernel Interpolation for Scalable Structured Gaussian Processes (KISS-GP)**
The kernel of GP implicitly depends on the kernel hyperparameters, such as the length-scale, signal variance, and noise variance (collectively denoted as θ), which are unknown and are inferred from the data. The fully Bayesian posterior inference of θ is non-trivial and often intractable. Hence, to avoid complexity, the standard practice is to obtain point estimates of θ by maximizing the marginal log-likelihood as

$$\log p(\mathbf{y}|\theta) \propto -[\mathbf{y}^T (K_\theta + \sigma^2 I)^{-1} \mathbf{y} + \log |K_\theta + \sigma^2 I|] \quad (9)$$

However, evaluation of $(K_\theta + \sigma^2 I)^{-1} \mathbf{y}$ and $K_\theta + \sigma^2 I$ require $O(n^3)$ and $O(n^2)$ operations, respectively.

Approaches like the subset of regressors (SoR)[18] and fully independent training conditional (FITC)[19] are the traditional approaches that are used to scale the GP inference. Recently, Wilson *et al.*[16] introduced a structured kernel interpolation (SKI) framework, which generalizes point estimate methods such as FITC and SoR for scalable GP inference. For instance, the kernel in SoR approach, $k_{SOR}$, is computed as

$$k_{SOR}(x, x') = k_{xU} K_{UU}^{-1} K_{Ux^T} \tag{10}$$

where $k_{xU}(size\ 1\ \times\ n), K_{UU}^{-1}(size\ m\ \times\ m), K_{Ux^T}(size\ n\ \times\ 1)$ are covariance matrices generated from the exact kernel $k(x, x')$ for a set of $m$ inducing points $[u_1, ..., u_m]$. Under the SKI framework, the exact kernel is replaced to an approximate kernel for fast computation by modifying $k_{SOR}$ considering $k_{xU} \approx W K_{UU}$, where W is an $n \times m$ matrix of interpolation, which is extremely sparse. Therefore Equation (10) can be rewritten as

$$k_{SOR}(x, x') \approx K_{xU} K_{UU}^{-1} K_{Ux^T} \approx W K_{UU} K_{UU}^{-1} K_{UU} W^T = W K_{UU} W^T = K_{SKI} \tag{11}$$

This general approach to approximating GP kernel functions is the basic framework of SKI[40], which in turn reduces the computation expense considerably, up to $O(n)$.

**Massively Scalable Gaussian Process (MSGP)**

While KISS-GP make learning faster up to $O(n)$, test predictions computational complexity is the same as in the traditional GP. Wilson *et al*[17] introduced MSGP, which extends KISS-GP to: (i) make test predictions significantly faster up to $O(1)$, (ii) scale marginal likelihood evaluations without requiring any grid structure, and (iii) project input data to lower dimensional space to avoid the curse of dimensionality. In MSGP predictions, the predictive mean is computed as

$$\text{mean}(f_*) \approx W_* K(U, U) W^T (K(X, X) + \sigma_\epsilon^2 I)^{-1} y \tag{12}$$

This is done by approximating $K(X_*, X)$ employing SKI as given by Equation (11)[36]. Here, we have to pay attention to the fact that the term $K(U, U) W^T (K(X, X) + \sigma_\epsilon^2 I)^{-1} y$ is pre-computed during training reducing the cost of online computations to $O(1)$. In similar lines, predictive covariance is computed as

$$\text{covariance}(f_*) \approx \text{diag}(K(X_*, X_*)) - \text{diag}(\{K(X_*, X)(K(X, X) + \sigma_\epsilon^2 I)^{-1} K(X, X_*)\}) \tag{13}$$

The diagonal operator in Equation (13) is the consequence of the fact the kernel matrices are highly sparse in the non-diagonal directions. Covariance computations in Equation (13) can be further simplified utilizing SKI as follows

$$\text{covariance}(f_*) \approx \text{diag}(K(X_*, X_*)) - W_* \text{diag}(\{K(U, X)(K(X, X) + \sigma_\epsilon^2 I)^{-1} K(X, U)\}) \tag{14}$$

Here, the term $\text{diag}(K(U, X)(K(X, X) + \sigma_\epsilon^2 I)^{-1} K(X, U))$ can also be pre-computed[36], leading to the overall computational cost of evaluating the predictions reducing to $O(1)$.

**Model training and hyperparametric optimization**

The model training for GPR was carried out using the GPytorch[20] python library. We optimize the GPR models considering all the following objectives: (i) training loss is minimum, (ii) mean squared error (MSE) for the full dataset (that is, MSE (training) + MSE (test)) is minimum, and (iii) training $R^2$ should be comparable to test $R^2$. Five-fold cross-validation was implemented with 60:20:20 for the training, validation, and test set, to avoid overfitting. Note that the test set was used only at the end for the model evaluation, while the model selection was carried out using the training and validation set. Further, hyperparametric optimization was carried out by varying (i) learning rate, (ii) weight decay, and (iii) considering various kernel functions. We observed that the radial basis functions provided the best predictions. Hence, this kernel function was implemented for all the properties.

**Results and Discussion**

Figure 1 shows the distribution of the nine properties in the processed dataset used for training the GPR models. We observe that all the properties in the dataset are distributed over a wide range, most of them spanning over an order of magnitude. Note that an exhaustive dataset cleaning and preparation were performed on the raw dataset (see Methods and Ref. [14]). Precisely, the final dataset consists of 37 components, namely, $SiO_2$, $B_2O_3$, $Al_2O_3$, MgO, CaO, BaO, $Li_2O$, $Na_2O$, $K_2O$, $Ag_2O$, $Cs_2O$, $Ti_2O$, BeO, NiO, CuO, ZnO, CdO, PbO, $Ga_2O_3$, $Y_2O_3$, $La_2O_3$, $Gd_2O_3$, $Bi_2O_3$, $TiO_2$, $ZrO_2$, $TeO_2$, $P_2O_5$, $V_2O_5$, $Nb_2O_5$, $Ta_2O_5$, $MoO_3$, $WO_3$, $H_2O$, $Sm_2O_3$, $MgF_2$, $PbF_2$, and $PbCl_2$, and nine properties, namely, density, Young's, shear, and bulk moduli, TEC, Vickers' hardness, refractive index, $T_g$, and liquidus temperature. These represent the most extensive composition–property models developed in the glass science literature covering most of the man-made glass compositions[14]. Further details on the dataset, including the distribution of the glass compositions with respect to number components and for each of the input components, are provided in the supplementary material (see Figs S1 and S2).

We train this dataset employing KISS-GP (see Methods) with hyperparametric tuning to develop optimized models. Note that we use KISS-GP[16] with Lanczos variance estimates (LOVE)[21], which significantly reduces the computational time and storage complexity (see Methods). Further, the prediction for high-dimensional data is carried out using MSGP. Due to the $\mathcal{O}(1)$ nature of MSGP[17], computational resources associated with the prediction are independent of the size of the data, thus enabling faster predictions (see Methods for details). Figure 2 shows the predicted values of density, Young's, shear, and bulk moduli, TEC, Vickers' hardness, refractive index, $T_g$, and liquidus temperature, in comparison to the measured experimental values for the trained GPR model with KISS-GP and MSGP. Since there are significant overlapping points, a heat map is used in Figure 2, wherein the number of points associated with each property per unit area is represented by the respective coloring scheme. The inset related to each property shows the probability density of error in the prediction with a confidence interval of 90%. We observe that the $R^2$ values for all the properties are equal to or greater than 0.8, suggesting a well-trained model. Further, the $R^2$ values of the training, validation, and test set are comparable, thereby confirming the goodness-of-fit of the model.

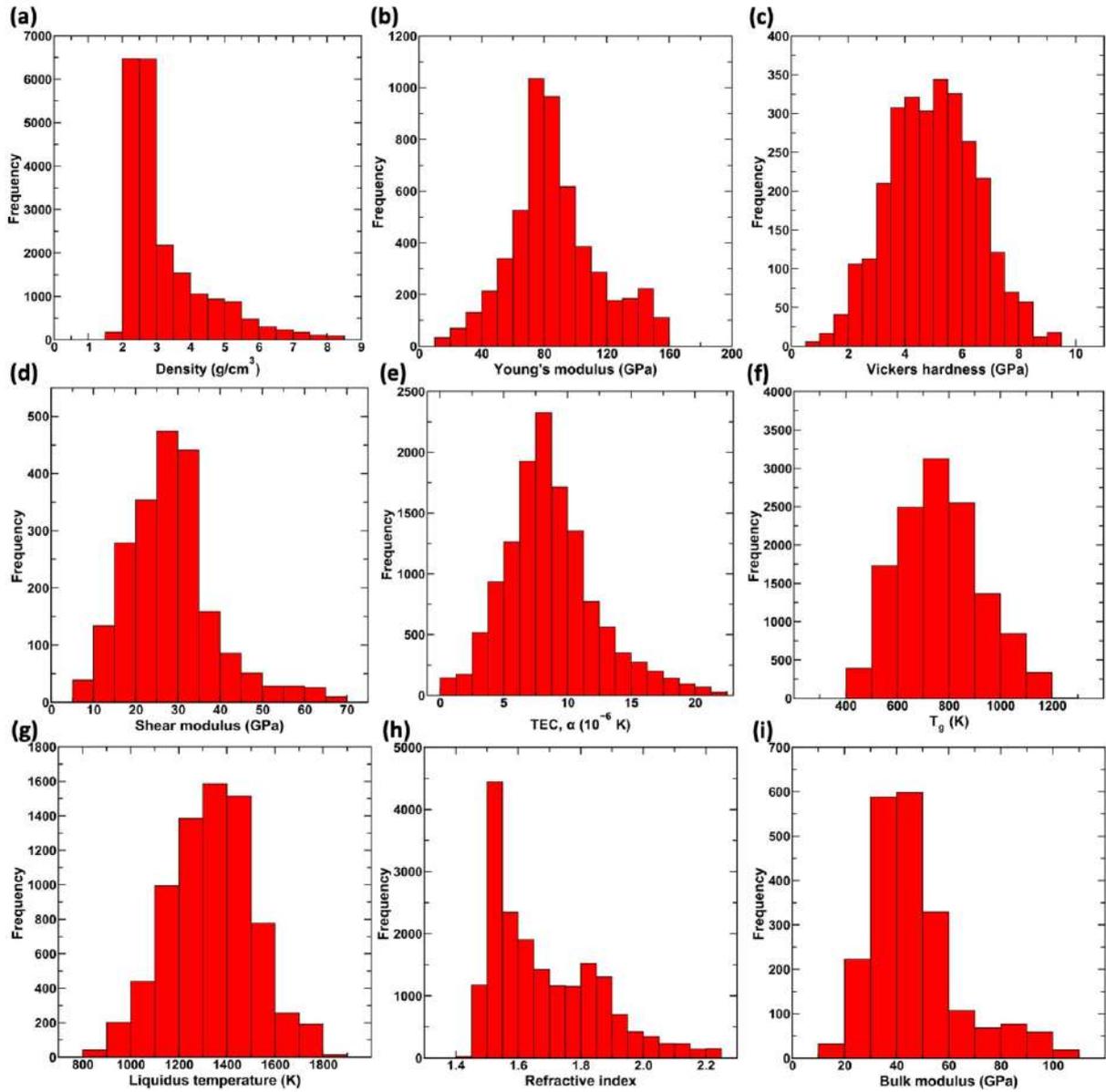

**Figure 1. Dataset visualization.** Distribution of the properties, namely, (a) density, (b) Young's modulus, (c) hardness, (d) shear modulus, (e) thermal expansion coefficient (TEC), (f) glass transition temperature ($T_g$), (g) liquidus temperature, (h) refractive index, and (i) bulk modulus, in the glass dataset used for training the GPR model.

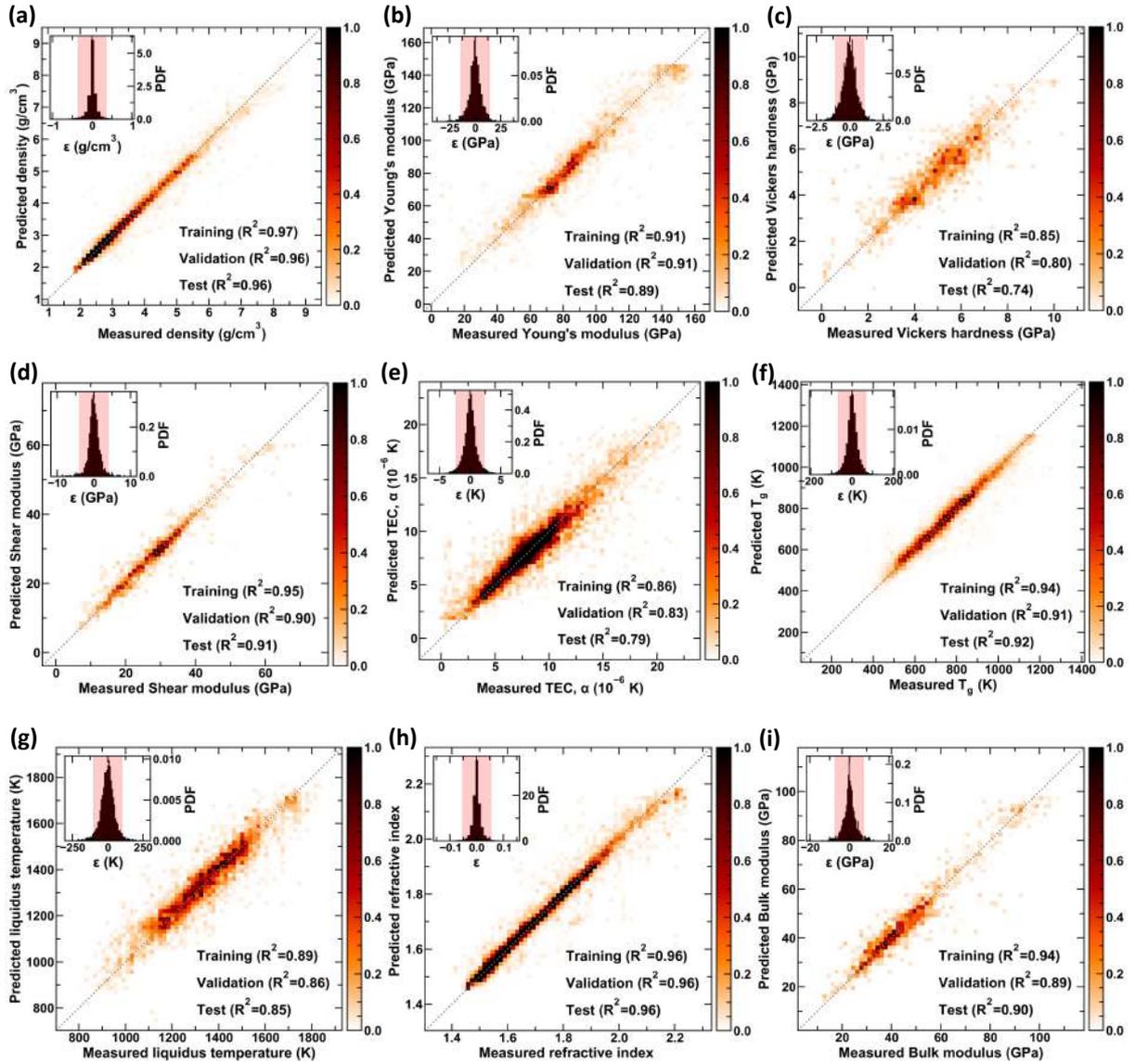

**Figure 2. KISS-GP models for the prediction of properties.** Predicted values of (a) density, (b) Young's modulus, (c) hardness, (d) shear modulus, (e) thermal expansion coefficient (TEC), (f) glass transition temperature ($T_g$), (g) liquidus temperature, (h) refractive index, and (i) bulk modulus using GPR with respect to the experimental values. Due to a large number of overlapping points, the color scheme is used to represent the number of points per unit area associated with each. INSET shows the error in the predicted values as a probability density function (PDF) with the shaded region representing the 90% confidence interval. Training, validation, and test $R^2$ values are also provided separately.

Now, we demonstrate one of the key attractive features of the proposed GPR-based approach to provide the uncertainty associated with the prediction. To this extent, we choose a ternary glass composition of sodium borosilicate, $x$(Na$_2$O).$y$(B$_2$O$_3$).1-$x$-$y$(SiO$_2$). Figure 3 shows the standard deviation of predicted values for the entire range of the ternary using the trained KISS-GP model. The mean values of the properties representing the best estimate of the model for this ternary are provided in the Supplementary material). The compositions corresponding to the measured values in the original data (which may belong training, validation, or test set) are marked using the black squares. We observe that the standard deviation in the predictions of compositions close to the original dataset is significantly low. As the compositions are farther away from from the original dataset, the standard deviation of the predicted value increases.

This behavior is consistently observed for all the properties (see Fig. 3). This is because, in KISS-GP, the training is carried out by identifying the distribution that reduces the variance for a known data point (that is training data) to zero, or atleast very close to it. As such, when the model is extrapolated to domains without any training data, the inference becomes poor as represented by larger standard deviation values. Nevertheless, we observe that the standard deviation for most of the properties are relatively low confirming a high confidence in the values predicted by the model. Overall, we observe that KISS-GP allows the development of reliable composition–property models, quantifying uncertainty in predictions when extrapolated over the entire compositional space.

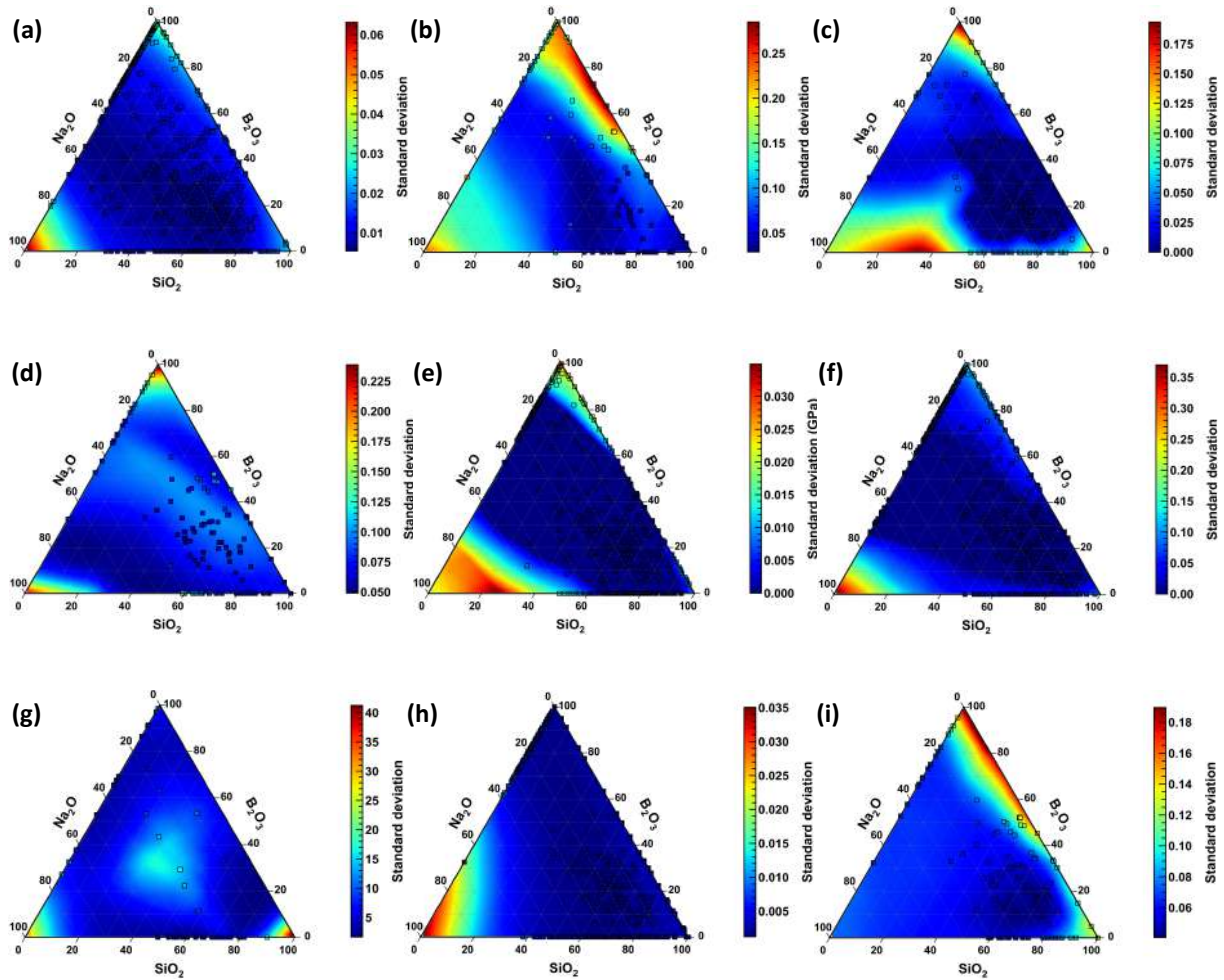

**Figure 3. Standard deviation for predicted values using KISS-GP.** Standard deviation predicted by the trained GPR models of (a) density, (b) Young's modulus, (c) hardness, (d) shear modulus, (e) thermal expansion coefficient (TEC), (f) glass transition temperature ($T_g$), (g) liquidus temperature, (h) refractive index, and (i) bulk modulus for sodium borosilicate glasses. Experimental values are marked using back squares.

| **Table 1.** Comparison of the $R^2$ values of the predictions KISS-GP with respect to linear regression, XGBoost, and deep neural networks (DNN)[14]. | | | | |
|---|---|---|---|---|
| **Property** | **Linear regression**[14] | **XGBoost**[14] | **DNN**[14] | **KISS-GP** |
| Density | 0.93 | 0.95 | 0.95 | **0.97** |
| Young's modulus | 0.79 | 0.87 | 0.86 | **0.90** |
| Hardness | 0.62 | 0.77 | 0.80 | **0.80** |
| Shear modulus | 0.78 | 0.88 | 0.88 | **0.92** |
| Thermal expansion coefficient | 0.69 | 0.80 | 0.82 | **0.83** |
| Glass transition temperature | 0.80 | 0.89 | 0.90 | **0.92** |
| Liquidus temperature | 0.60 | 0.82 | 0.80 | **0.85** |
| Refractive index | 0.92 | 0.94 | 0.95 | **0.96** |
| Bulk modulus | 0.75 | 0.89 | 0.89 | **0.90** |

Now, we compare the performance of the GPR models with some of the state-of-the-art ML models[14]. Table 1 shows the $R^2$ values of KISS-GP in comparison to linear regression, XGBoost, and DNN (see Ref.[14] for details). Note that the dataset used for training all the models are the same[14]. Interestingly, we observe that the goodness-of-fit of KISS-GP represented by the $R^2$ values is superior to all other methods. This trend is consistently observed for all nine properties. This result confirms that the predictions obtained from KISS-GP and MSGP are reliable and superior to other state-of-the-art methods as presented in Table 1. Besides, the uncertainty quantification in KISS-GP and MSGP-based property predictions is quite useful to interpret the model validity in experimentally unexplored regimes of the compositional space, a feature that is severely lacking in the other deterministic models.

**Conclusion**

Overall, employing KISS-GP and MSGP, we show that reliable composition–property models can be developed for large datasets. These models for predicting density, Young's, shear, and bulk moduli, TEC, Vickers' hardness, refractive index, $T_g$, and liquidus temperature of inorganic glasses with up to 37 input components, the largest so far, allows the exploration of a large compositional space that was hitherto unknown. The KISS-GP models, in addition to being able to quantify uncertainty, yield superior results when compared against state-of-the-art methods such as XGBoost or DNN models. The generic approach presented here can be applied for developing composition–property relationships of a wide range of materials involving large data with detailed uncertainty quantifications. The models developed here have been made available as part of a package, Python for Glass Genomics (PyGGi, see: http://pyggi.iitd.ac.in)[22].

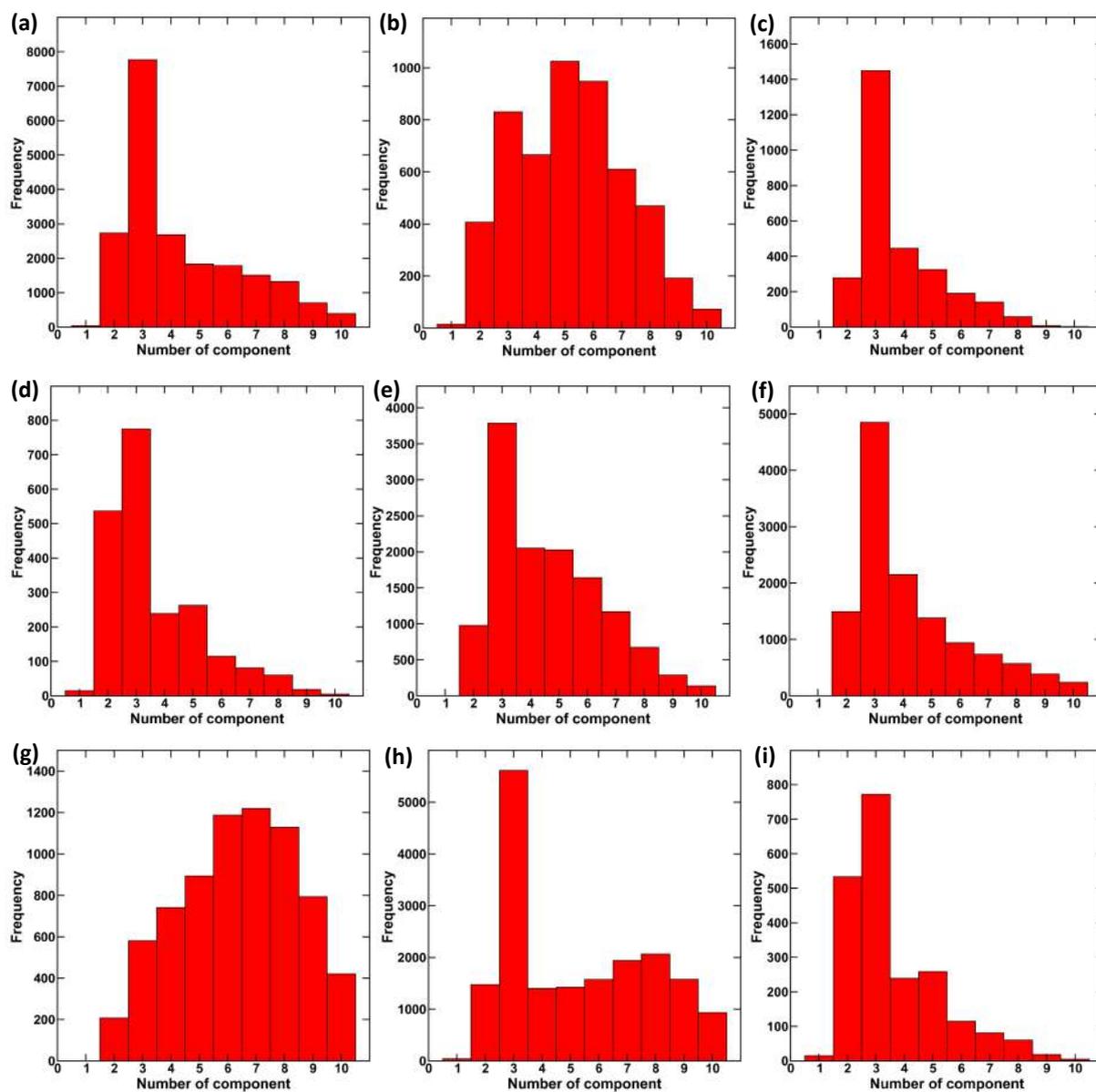

**Figure S1.** Frequency of glasses with respect to their number of components for (a) density, (b) Young's modulus, (c) hardness, (d) shear modulus, (e) thermal expansion coefficient (TEC), (f) glass transition temperature ($T_g$), (g) liquidus temperature, (h) refractive index, and (i) bulk modulus.

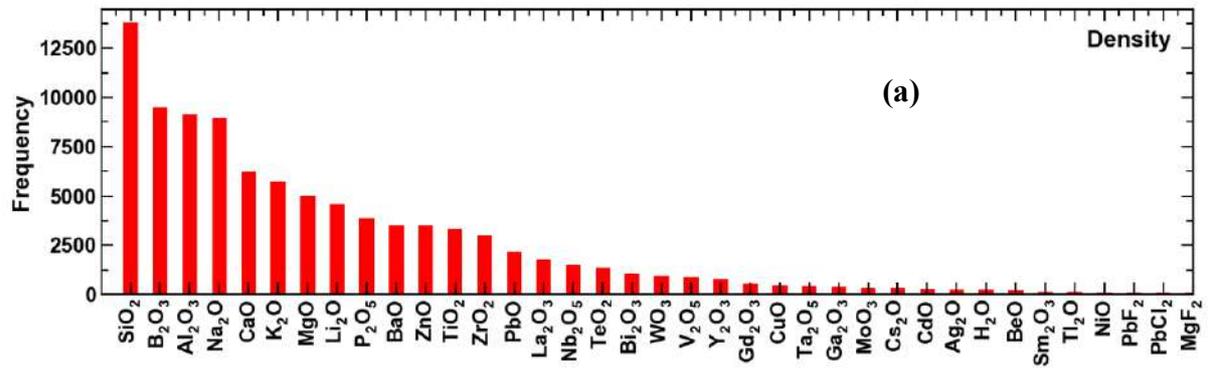
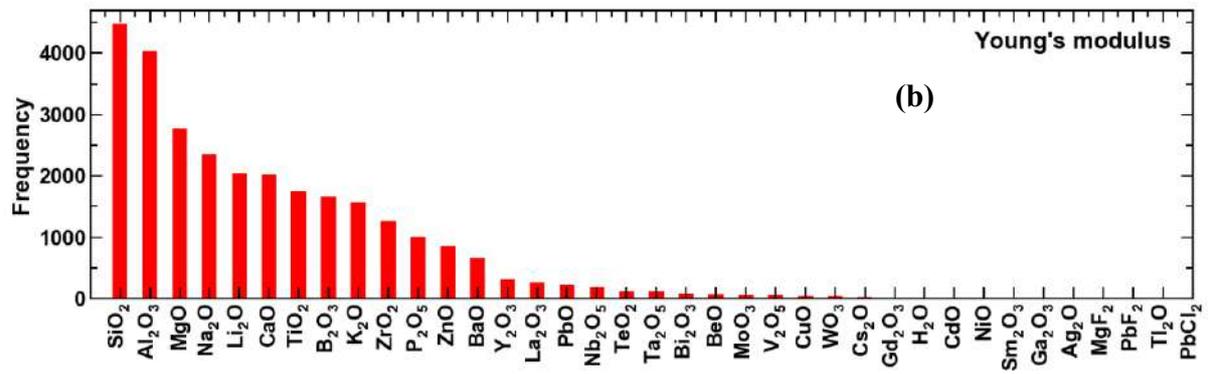
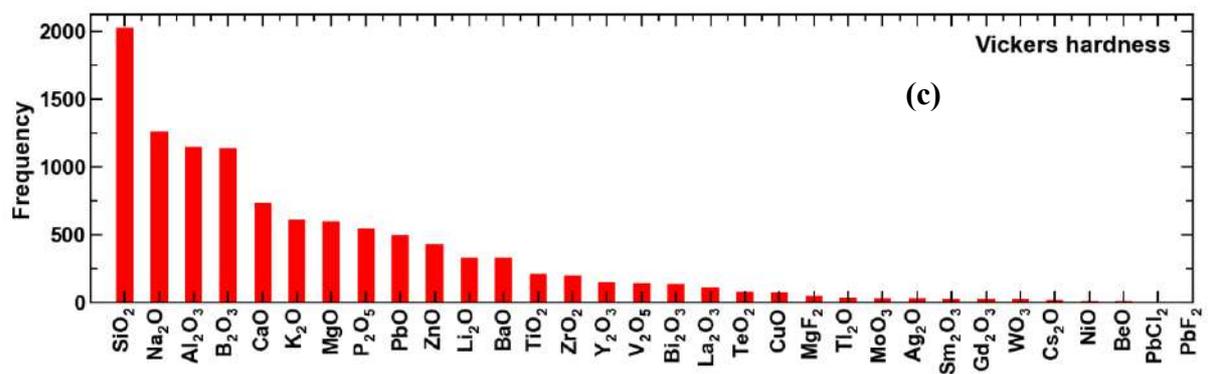
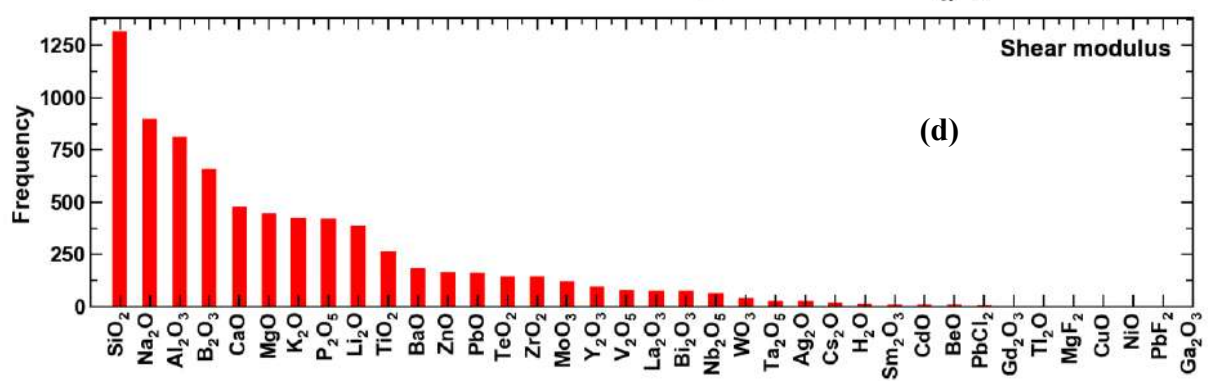

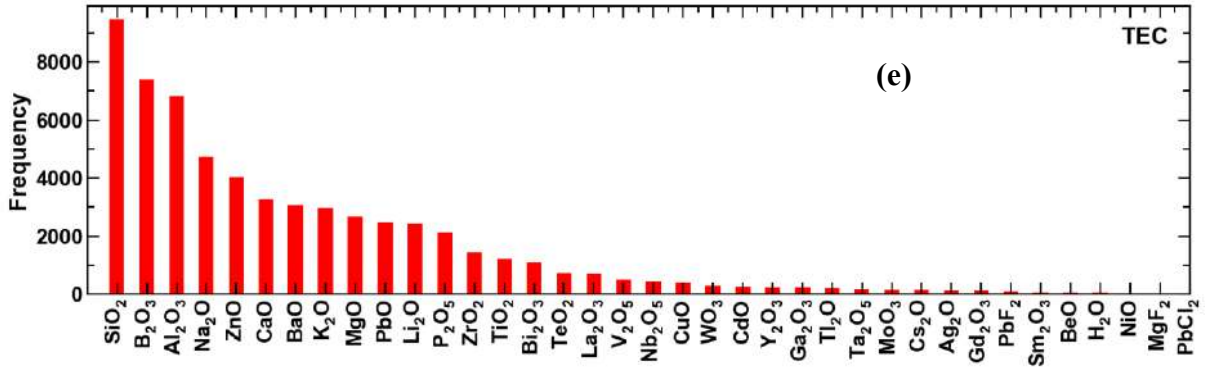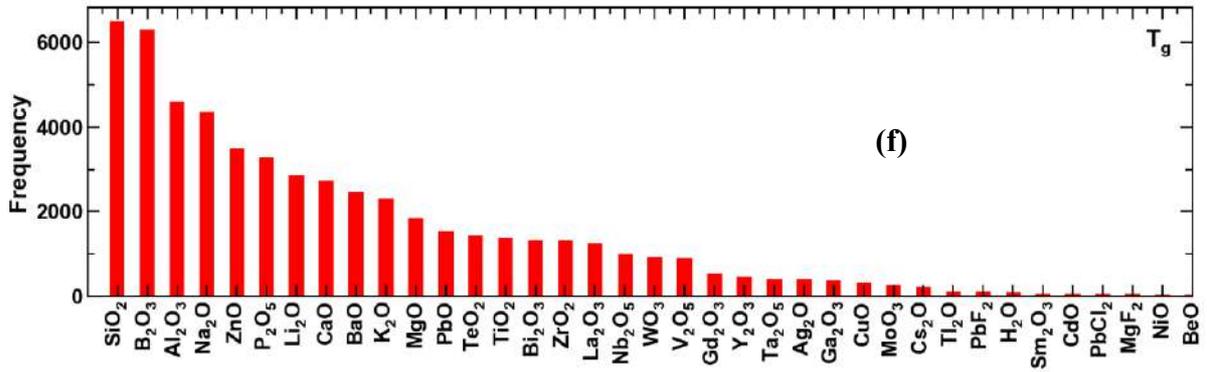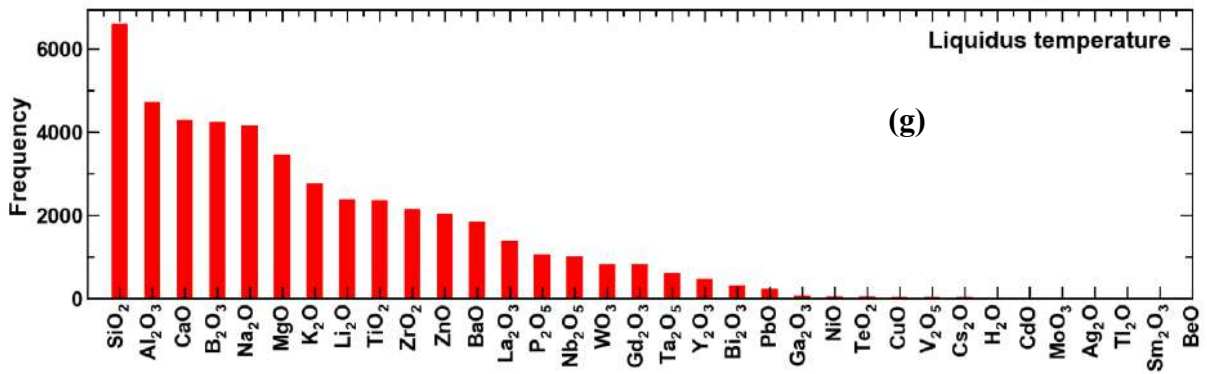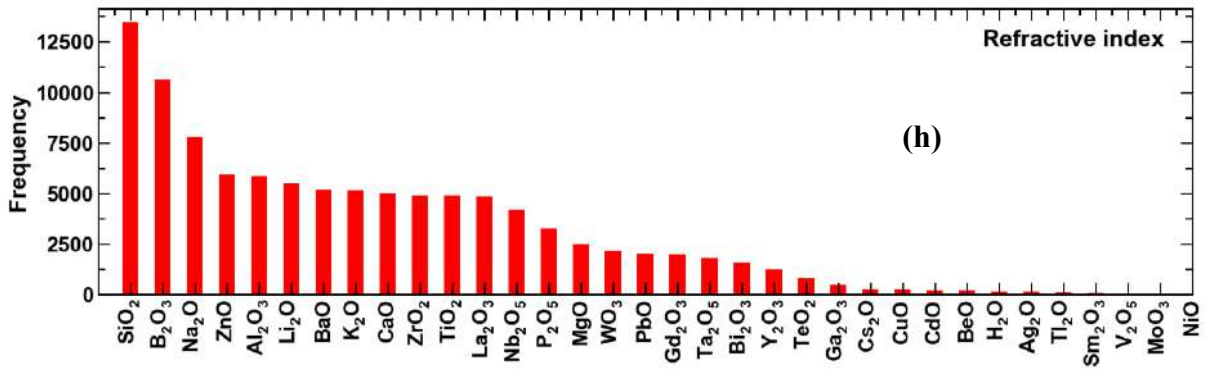

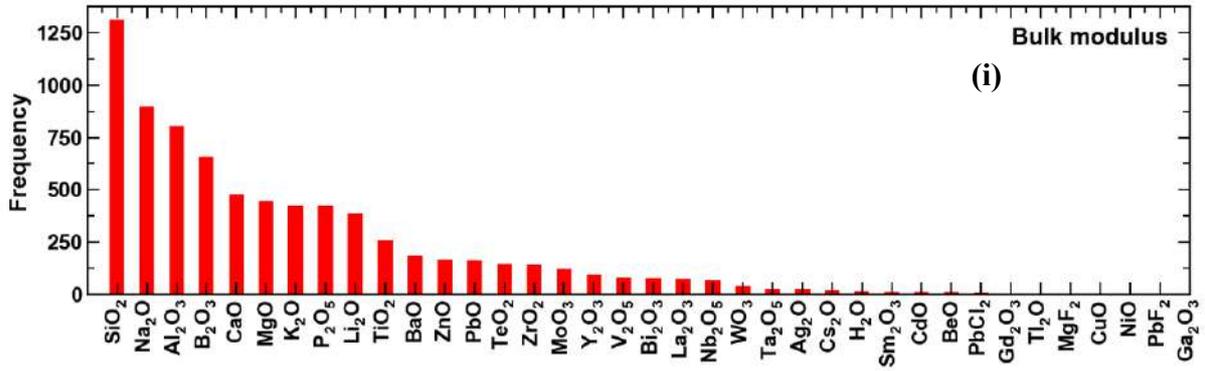

**Figure S2.** Frequency of glasses with respect to each of the input components for (a) density, (b) Young's modulus, (c) hardness, (d) shear modulus, (e) thermal expansion coefficient (TEC), (f) glass transition temperature ($T_g$), (g) liquidus temperature, (h) refractive index, and (i) bulk modulus.

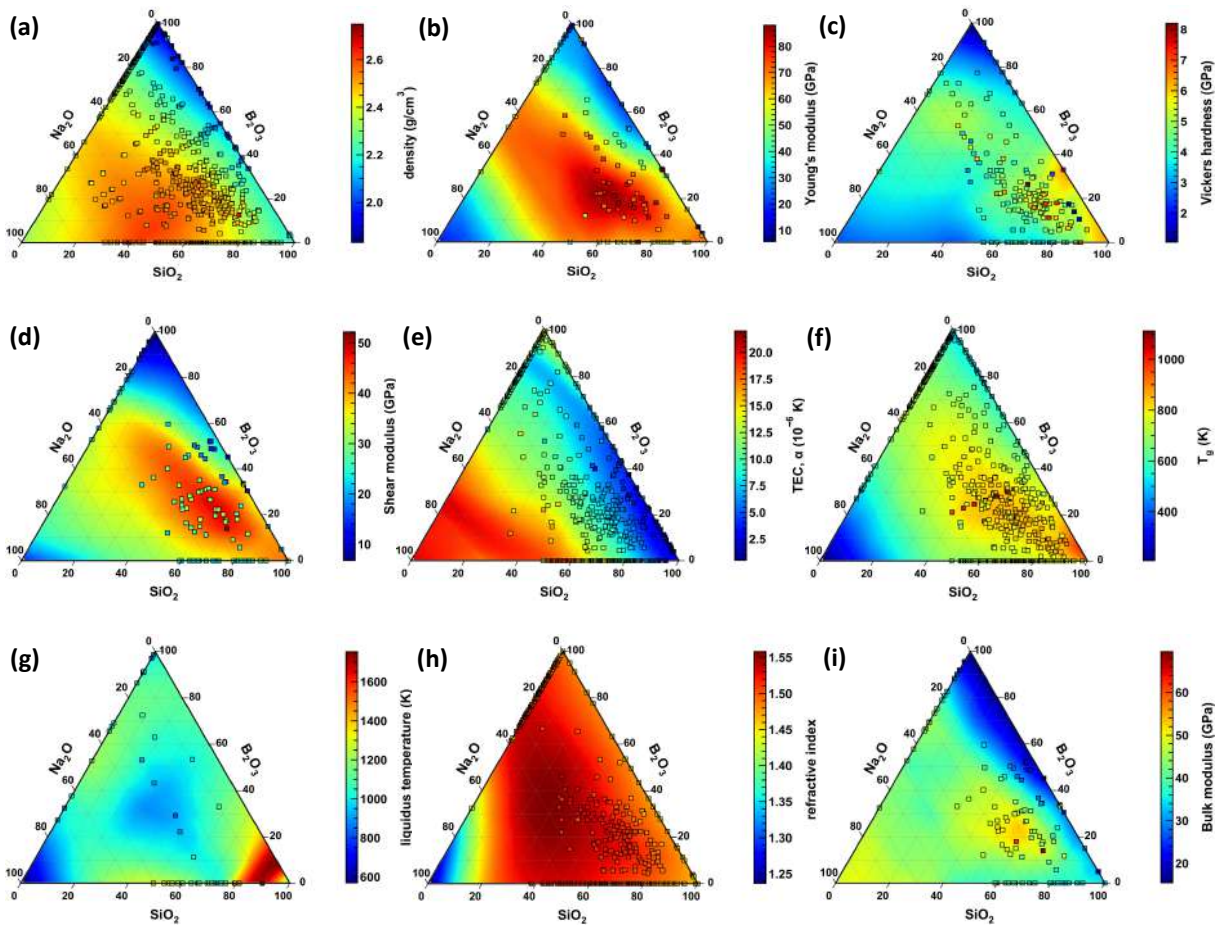

**Figure S3.** Predicted values of (a) density, (b) Young's modulus, (c) hardness, (d) shear modulus, (e) thermal expansion coefficient (TEC), (f) glass transition temperature ($T_g$), (g) liquidus temperature, (h) refractive index, and (i) bulk modulus by the trained GPR models for sodium borosilicate glasses. Experimental values are marked using back squares.